\documentclass[preprint]{aastex}

\shorttitle{X-ray slope-Eddington ratio correlation in quasars}
\shortauthors{Risaliti et al.}

\begin{document}

\title{The SDSS/XMM-Newton Quasar Survey: \\
    Correlation between X-ray spectral slope and Eddington ratio}

\author{G. Risaliti\altaffilmark{1,2}, M. Young\altaffilmark{1,3}, and M.~Elvis\altaffilmark{1}}

\altaffiltext{1}{Harvard-Smithsonian Center for Astrophysics,
    60 Garden Street, Cambridge, MA 02138, USA}
\altaffiltext{2}{INAF-Osservatorio Astrofisico di Arcetri, Largo E. Fermi 5, I-50125 Firenze, Italy}
\altaffiltext{3}{Boston University, Astronomy Department, 725 Commonwealth Ave., Boston, MA 02215, USA}

\begin{abstract}
We present a correlation between the 2-10~keV spectral slope $\Gamma_X$ and the
Eddington ratio L/L$_{EDD}$ in a sample of $\sim400$ Sloan Digital Sky Survey quasars 
with available hard X-ray spectra from {\em XMM-Newton} serendipitous observations.
We find that the $\Gamma_X$-L/L$_{EDD}$ correlation is strongest in objects with black hole (BH)  masses 
determined from the $H\beta$ line, and weaker (but still present) for those
based on Mg~II. An empirical non-linear correction of the
Mg~II-based masses, obtained by comparing the mass estimates in SDSS quasars having
both H$\beta$ and Mg~II measurements, significantly increases the strength of 
the correlation.  No correlation is found among objects with BH masses derived
from C~IV, confirming that this line is not a reliable indicator of the BH mass. 
No significant correlation is found with the bolometric luminosity,
while a $\Gamma_X$-M$_{BH}$ relation is present, though with a lower statistical  significance than 
between $\Gamma_X$ and L/L$_{EDD}$. Our results imply a physical link between the accretion
efficiency in the (cold) 
accretion disc of AGNs and the physical status of the (hot) corona responsible for the X-ray emission.
\end{abstract}





\section{Introduction}
The origin of the X-ray emission in quasars is not 
well understood. In the widely accepted disk-corona
model (Haardt \& Maraschi~1993), 
the X-rays are produced in a hot phase (the corona)
reprocessing the primary optical/UV emission of the disk.
However, the mechanisms of energy transfer to the hot phase,
and its geometry and size are not clear. Therefore, it
is not known how the basic physical parameters
(such as black hole mass, accretion rate, total luminosity)
affect the X-ray emission.
Laor et al.~(1997) found a correlation between the soft X-ray (0.2-2~keV) slope
and the full width at half maximum (FWHM) of the $H\beta$ emission
line in a sample of 23 low redshift quasars, and suggested that
the physical parameter driving the correlation is the Eddington ratio, 
L/$L_{EDD}$, where L is the bolometric luminosity. Further studies in the 2-10~keV energy
band (e.g. Brandt et al.~1997, Shemmer et al.~2006) 
confirmed this correlation for the hard X-ray slope, and is therefore not related to the
AGN ``soft excess''. In particular, a $\Gamma_X$-L/L$_{EDD}$ correlation has been found in a sample of 
$\sim150$ SDSS quasars with {\em Chandra} spectra, and spanning a large redshift
range (Kelly et al.~2008). 
Recently Shemmer et al.~(2008) presented a similar analysis based on
 a sample of 35 quasars,
spanning more than three orders of magnitude in luminosity, and 
obtained a stronger correlation between $\Gamma_X$ and the Eddington
ratio L/L$_{EDD}$ than with FWHM($H\beta$), breaking the partial degeneracy 
between these two quantities.
The main limitation of past studies involving X-ray slopes is that either
they rely on low quality X-ray data, or they are based on relatively small samples.
The  availability of large, homogeneous samples of quasars with 
high quality optical and X-ray spectral data is now opening new
possibilities in this field. 

In particular,
we recently obtained a new sample of $\sim$800 quasars within
the SDSS/{\em XMM-Newton} survey (Risaliti \& Elvis~2005, Young, Elvis \& Risaliti~2009,
hereafter Y09)
obtained  by cross-correlating
the SDSS DR5 quasar catalog with the {\em XMM-Newton} public archive. 
Since we only selected 
serendipitous {\em XMM-Newton} observations, this sample can
be considered as randomly extracted from the SDSS quasars.
About 500 quasars in this sample have good enough X-ray data to
perform a basic spectral analysis and obtain a continuum slope.
A complete description of the survey, with an analysis of the general
properties of the sample, is presented in Y09.

The simultaneous availability of optical/UV and X-ray spectra
not only allows a more precise analysis of the $\alpha_{OX}$-luminosity
correlation (this subject is developed in a companion
paper, Young, Risaliti \& Elvis~2009, in prep.), but also opens a whole new field of analysis of the possible
correlations among X-ray spectral parameters, optical/UV continuum and line emission, and
physical parameters of the AGN. 

Here we extend the analysis of the correlation between $\Gamma_X$ and
optically-derived and global quantities such as M$_{BH}$, L/L$_{EDD}$, and L
for our sample of SDSS-{\em XMM-Newton} quasars, consisting of more than 400 objects.

\section{The sample}

Our sample is taken from the SDSS/{\em XMM-Newton} quasar survey (Y09). Since we are interested
in a homogeneous analysis of X-ray slopes, we excluded radio loud (RL) quasars and 
broad absorption line (BAL) quasars, whose X-ray spectra may be contaminated by synchrotron emission (in RLs)
or affected by heavy absorption (in BALs). Since a complete removal of BALs is possible only for sources at z$>$1.5, and $\sim$half of our sources have z$<$1.5, there may be some residual BALs contamination in our sample. By removing objects with flat X-ray spectra (see below) we expect to further clean the sample from BALs, which are typically harder in X-rays than non-BAL quasars (e.g. Gallagher et al.~2006). The remaining spurious objects are expected to be $\lesssim$5\% of the sample.  

For the 403 quasars with a signal-to-noise ratio S/N$>$6, we used the available X-ray and optical/UV spectral data 
to obtain estimates of the relevant physical parameters for our analysis:
the X-ray continuum slope, $\Gamma_X$, the black hole mass, $M_{BH}$, and
the bolometric luminosity, L$_{BOL}$.
The sample spans about three orders of magnitude in optical luminosity (this can be estimated 
from the bottom panel of Fig.~1, where we
show the bolometric luminosities, which are roughly proportional to the optical ones, as explained below). 
The 2-10~keV luminosities range from 10$^{43}$ to 10$^{45.5}$~erg~s$^{-1}$ for most sources, with small tails down to 10$^{42}$~erg~s$^{-1}$ at z$<$0.5, and up to 10$^{46}$~erg~s$^{-1}$ at z$>$1.5. More details on the luminosity distribution can be found in Y09.

The X-ray slopes $\Gamma_X$ were obtained from 
basic power law plus Galactic absorption models applied to the spectral data in the rest-frame 2-10~keV band. 
The data reduction was performed with the SAS package\footnote{http://xmm.esac.esa.int/sas/8.0.0/} 
following the standard recommended steps
in the {\em XMM-Newton} Science Data Center web page. The analysis was made with the {\em Sherpa} analysis
package\footnote{http://cxc.harvard.edu/sherpa/}. All the details about the data reduction and analysis are presented in Y09.
The sample spans an X-ray luminosity $10^{43}$~erg~s$^{-1}<L_X<
5\times10^{46}$~erg~s$^{-1}$, and a redshift range 0.1$<$z$<$4.5.

\begin{figure}[h!]
\epsscale{0.6}
\plotone{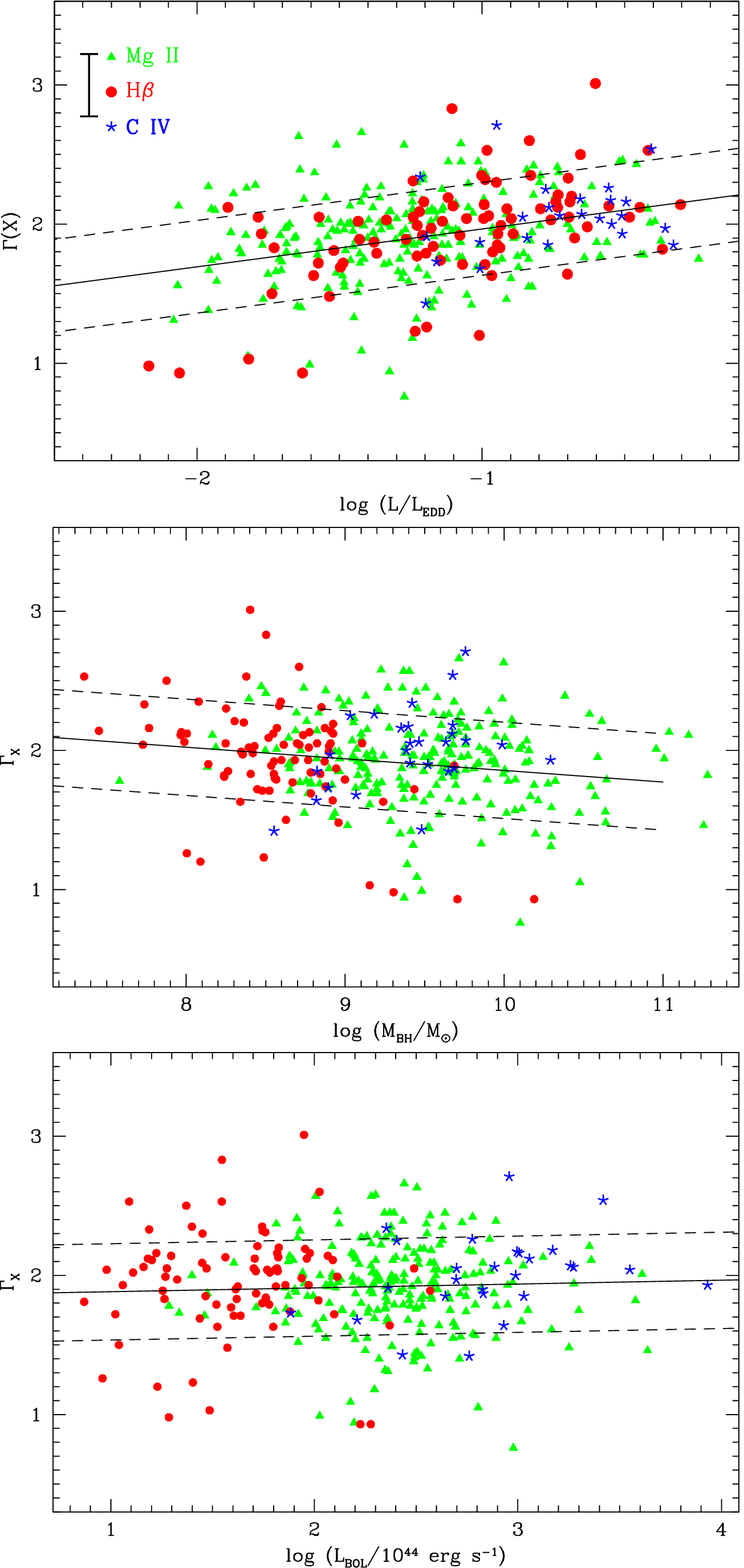}
\figcaption{\footnotesize{Correlations between $\Gamma_X$ and L/L$_{EDD}$, $M_{BH}$, and $L_{BOL}$
for our sample. The error bars are not shown for clarity. A typical error is shown in the 
top panel. The three colors refer to the line used to estimate the black hole mass.
The lines show the best linear fits and the dispersions. The statistical errors on the slopes
are small with respect to the dispersions, as shown in Table~1, and are not plotted here.
}}
\label{totfit}
\end{figure}

The black hole masses have been estimated (Shen et al.~2008, hereafter S08) 
from the widths of the optical broad emission lines and
the underlying continuum luminosity.
Three different optical lines are used, depending on the redshift of the sources: $H\beta$ ($0<z<0.9$),
Mg~II ($0.4<z<2.2$), and C~IV ($1.7<z<4.5$). 
When the black hole masses from two different lines are available, 
we prefer $H\beta$ over Mg~II, and Mg~II over C~IV (see Section~4 for more details on this issue).
For the H$\beta$ and Mg~II lines,
the widths were obtained through a two-Gaussian fit, after subtracting 
a template reproducing the Fe~II emission.
The adopted templates are those of Boroson \& Green~(1992) for
the H$\beta$ line and of Salviander et al.~(2007) for the Mg~II line. For the C~IV line,
the continuum has been fitted with a simple power law, while the line FWHM has been
estimated from the analytic three-Gaussian fit described in 
Laor et al.~(1994). Full details on the optical spectral analysis are
described in S08.

The bolometric luminosities are derived from the continuum monochromatic luminosity at the  line 
wavelength, adopting a correction based on an analytical intrinsic
Spectral Energy Distribution (SED). The optical to UV SED is approximated by three
power laws with spectral indexes (in a $\nu-f_\nu$ plane) $\alpha_1=-2$ in the 1-10~$\mu$m  
interval\footnote{This 
part gives a negligible contribution, and  represents the Raileigh-Jeans tail of the accretion disk
emission. The actual observed emission at wavelengths $\lambda>1~\mu$m is due to reprocessing by dust,
and therefore in not included here.}, 
$\alpha_2=-0.44$ from $1~\mu$m to 1,200~\AA, and $\alpha_3=-1.76$ from
1,200~\AA~to 500~\AA, in agreement with the average SED of SDSS quasars (Richards et al.~2006).
The X-ray luminosity is modeled with a power law continuum with the observed photon index, starting at 0.1~keV,
and an exponential decrease
with a cut-off energy $E_C=100$~keV. The X-ray to optical ratio is directly obtained from the observed
data.

We analyzed the correlation between the X-ray photon index, $\Gamma_X$, and: 
(1) the Eddington ratio, L/L$_{EDD}$, 
(2) the bolometric luminosity, 
$L_{BOL}$, and (3) the black hole mass, M$_{BH}$.
The main results for the whole sample are shown in Fig.~1. 
As mentioned above, our data consist of three different subsamples, depending on the
line used to estimate the black hole mass. 
There are at least two reasons to perform a separate analysis for each subsample: (1) the 
black hole mass-line width correlation is directly calibrated using reverberation mapping 
on $H\beta$ widths (e.g. Peterson et al.~2004),
while it is indirectly calibrated using $H\beta$ for the other lines (Vestergaard et al.~2006); 
(2) the luminosity and redshift
ranges spanned by the three lines are quite different (Fig.~1, bottom panel), 
and a possible luminosity dependence 
has to be considered.
We therefore repeated the same analysis for the total sample and each of the three subsamples. 
\begin{table*}
\caption{Results of the correlation analysis}
\centerline{\begin{tabular}{lcccccc}
Sample/Corr. & N$^a$ & P$^b$ & m$^c$ & q$^d$ & Disp$^e$& r$^f$ \\
\hline
Total: $\Gamma_X$-L/L$_{EDD}$    & 343 & $<10^{-8}$         & 0.31$\pm0.06$ & 1.97$\pm0.02$ & 0.33 & 0.32\\
$H\beta$: $\Gamma_X$-L/L$_{EDD}$ & 82  & $<10^{-8}$         & 0.58$\pm0.11$ & 1.99$\pm0.04$ & 0.35 & 0.56\\
$Mg II$: $\Gamma_X$-L/L$_{EDD}$  & 290 & 3$\times$10$^{-4}$ & 0.27$\pm0.09$ & 1.88$\pm0.03$ & 0.33 & 0.21\\
$Mg II$(corr)$^g$: $\Gamma_X$-L/L$_{EDD}$  & 290 & 3$\times$10$^{-6}$ & 0.24$\pm0.05$ & 1.98$\pm0.02$ & 0.32 & 0.30\\
$ C IV$: $\Gamma_X$-L/L$_{EDD}$  & 58  & 0.47               & -0.11$\pm0.17$ & 2.03$\pm0.06$ & 0.27 & -0.13\\
\hline
Total: $\Gamma_X$-M$_{BH}$     & 343 & 0.004              &-0.22$\pm0.08$ & 1.981$\pm0.03$& 0.35 &-0.16\\
$H\beta$: $\Gamma_X$-M$_{BH}$  & 82  & 0.003              &-0.33$\pm0.10$ & 1.95$\pm0.04$ & 0.39 &-0.35\\
$Mg II$: $\Gamma_X$-M$_{BH}$   & 290 & 6$\times$10$^{-4}$ &-0.11$\pm0.03$ & 2.01$\pm0.03$ & 0.33 &-0.21\\
$C IV$: $\Gamma_X$-M$_{BH}$    & 58  & 0.04               & 0.27$\pm0.10$ & 1.77$\pm0.08$ & 0.26 & 0.27\\
\hline
Total: $\Gamma_X-L_{BOL}$      & 343 & 0.45             & 0.03$\pm0.04$ & 1.90$\pm0.04$ & 0.35 & 0.04\\
$H\beta$: $\Gamma_X-L_{BOL}$   & 82  & 0.12             & 0.22$\pm0.15$ & 1.91$\pm0.05$ & 0.41 & 0.06\\
$Mg II$: $\Gamma_X-L_{BOL}$    & 290 & 0.10             &-0.05$\pm0.04$ & 2.02$\pm0.03$ & 0.34 &-0.07\\
$C IV$: $\Gamma_X-L_{BOL}$     & 58  & 0.06             & 0.23$\pm0.10$ & 1.70$\pm0.14$ & 0.27 & 0.20\\
\hline
\end{tabular}}
\footnotesize{$^a$: Number of object in each sample. 58 objects have both $H\beta$ and Mg~II measurements,
while 29 have both Mg~II and C~IV. $^b$: Probability of a null correlation from a Spearman's test.
$^c$,~$^d$: slope $m$ and intercept $q$ of a best fit linear correlation $Y=m\times(X-X_0)+q$.
In all the correlations $Y=\Gamma_X$. The independent variable $X$ and the reference point $X_0$ are:
X=log(L/L$_{EDD}$), $X_0=-1$ for the $\Gamma$-L/L$_{EDD}$ correlation, $X=log(M_{BH}/M_\odot)$, $X_0=8.5$
for the $\Gamma$-$M_{BH}$ correlation, and X=log(L$_{BOL}/10^{44}~{\rm erg~cm}^2~s^{-1}$), $X_0=1.5$
for the $\Gamma$-$L_{BOL}$ correlation. $^e$: Dispersion of the data points with respect to the 
linear correlation. $^f$: linear correlation coefficient. $^g$: Values obtained adopting the correction for
Mg~II based masses discussed in Section~3 and Fig.~3.
 }
\end{table*}
\begin{figure}[h!]
\epsscale{0.8}
\plotone{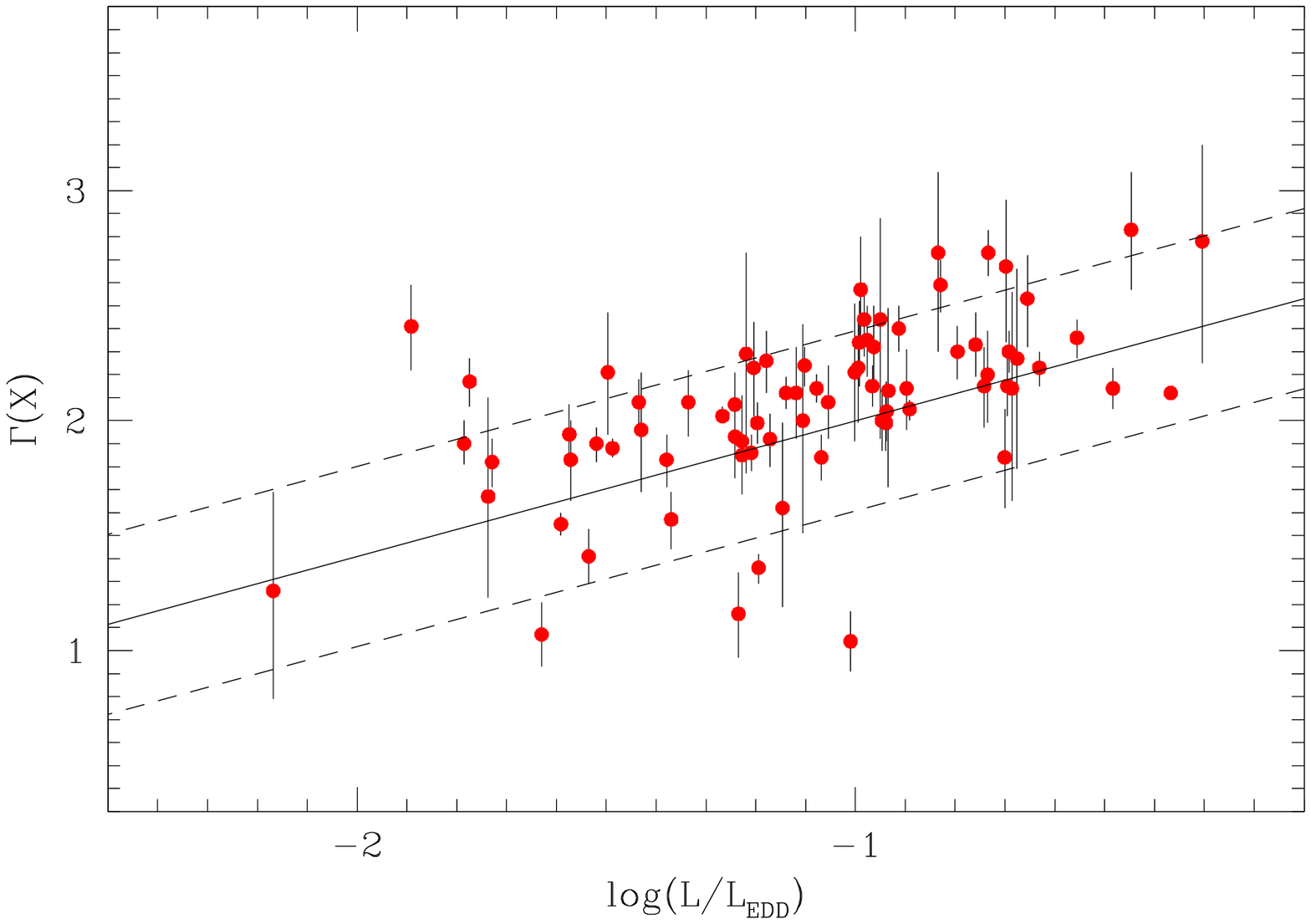}
\figcaption{\footnotesize{Correlations between $\Gamma_X$ and L/L$_{EDD}$ for the subsample of
objects with black hole mass determined from the $H\beta$ line. This is the strongest correlation
found in our analysis.}}
\label{totfit}
\end{figure}
\section{Statistical Analysis}
Our statistical analysis consists of several steps and checks.
We first performed a non-parametric Spearman rank test on each correlation, followed by a 
least squares linear fit. 
%
In order to take into account the effect of possible deviant points, we performed the
above analysis with different selections based on the quality of X-ray data:\\
(1) We tried different cuts
in signal-to-noise of the X-ray spectra (6$<$S/N$<$10), and in quality of the fit of the X-ray data
(0.5$<$$\chi_\nu^2$$<$2). 
No significant dependence on these choices has been found. The final analysis was performed
on objects with S/N$>$8 in the X-ray spectra (see Y09 for details)
and $\chi_\nu^2$$<$1.5 in the X-ray fit.\\
(2) We excluded possibly
``bad'' points, i.e. those with extreme best fit values of $\Gamma_X$ ($\Gamma_X$$>$3 or $\Gamma_X$$<$1), 
or large deviations from the best fit in a statistical sense (($\Gamma_X$-$<$$\Gamma_X$$>$)/$\Delta$$\Gamma_X$)$>$5.
Again, the results remain consistent for all the different cuts.
\\
(3) We estimated the errors for the slope $m$ and intercept $q$ through a bootstrap analysis, consisting
of repeating the linear fits on  randomly selected sets of data drawn from our sample, allowing for repetitions. 
This technique takes into account strongly deviating points that could affect the 
correlations.\\
All these checks demonstrate that our results are stable and do not depend on single deviating points.

The results for the total sample are shown in Table~1. The final sample  
 contains 343 objects.
The subsamples consist of 82 ($H\beta$), 290 (Mg~II) and 58 (C~IV) data points. 58 objects
have both $H\beta$ and Mg~II measurements, while 29 have both MG~II and C~IV.

\section{Results}
We found a highly statistically
significant correlation (probability of null correlation P$<0.1$\%) between the X-ray slope and the
Eddington ratio L/L$_{EDD}$, and a weaker, but still significant anti-correlation between $\Gamma_X$ and the
black hole mass.
No significant trend (P$>$5\%) is found with the
bolometric luminosity (Table~1). Separate analysis of the subsamples show that the $\Gamma_X$-L/L$_{EDD}$
correlation
is strongest for objects whose $M_{BH}$ is determined from $H\beta$ (Fig.~2), weaker for 
Mg~II objects, and absent for C~IV objects.
 The linear correlation coefficient is higher than 0.5, 
conventionally considered the threshold for a ``strong'' correlation, only  for the $\Gamma_X$-L/L$_{EDD}$ correlation
in the $H\beta$ group.

Since L/L$_{EDD}\propto M_{BH}^{-1}$, and 
$M_{BH}\propto$FWHM$^2 L^\alpha$ ($\alpha=$0.52$\pm$0.04 in the calibration of Bentz et al.~2006), 
a strong $\Gamma_X$-L/L$_{EDD}$ correlation 
may be due to an intrinsic correlation  between $\Gamma_X$ and
FWHM. 
We tested this possibility and found a highly significant $\Gamma_X$-FWHM correlation 
(slightly weaker than the $\Gamma_X$-L/L$_{EDD}$ one in the total and H$\beta$ samples).
As a more general approach, we tested correlations of the form
FWHM$^{-2}$$\times$ L$^\beta$ varying the exponent $\beta$ in the interval 0-0.8, 
and we always found the same level of correlation
for any value of $\beta$ 
(null correlation probability $P<10^{-8}$, linear correlation coefficient $r$ in the range 0.53-0.60).
This confirms that the luminosity term does not contribute to the observed correlation. This may
be due to either a physical independence of these quantities, or to a too narrow luminosity range
in our sample (Fig.~1). We note that Shemmer et al.~(2008) claim a stronger dependence of $\Gamma_X$ on
L/L$_{EDD}$ than on FWHM(H$\beta$) for their quasar sample, which is smaller (35 objects) but spanning a 
larger luminosity interval.

The $\Gamma_X$-log(M$_{BH}$) relation is statistically significant (Table~1), but rather weak, with a
slope of $\sim0.3$. Since $L_{EDD}$$\propto$M$_{BH}$, some degree of correlation of $\Gamma_X$ 
with $M_{BH}$ is
expected, given the strength of the $\Gamma_X$-L/L$_{EDD}$ correlation. 
Formally, it is not possible to remove this partial degeneracy.
We only note that the stronger $\Gamma_X$-L/L$_{EDD}$ correlation suggests that this is the physically
relevant relation, and, therefore, that there is no independent support for a direct physical dependence of  
$\Gamma_X$ on $M_{BH}$.
\begin{figure}[h!]
\epsscale{0.8}
\plotone{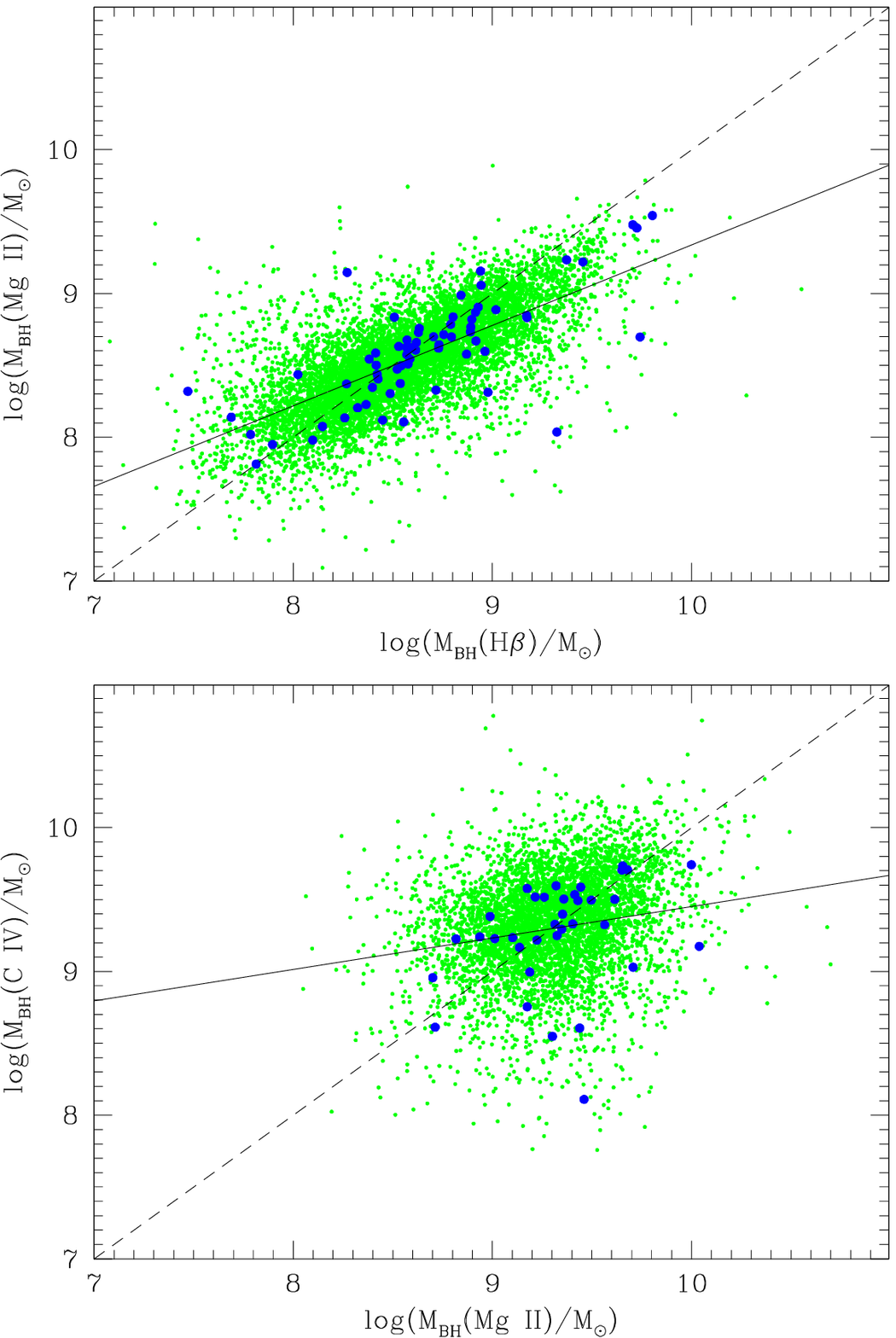}
\figcaption{\footnotesize{Comparison between the S08 
black hole mass estimates from different lines. The blue points are the objects in the SDSS/{\em XMM-Newton}
 sample. The continuous lines show the linear best fits (though the linear correlation in the
C~IV-Mg~II plane is barely significant, see text for details). The dashed lines 
indicate the one-to-one relation.}}
\label{totfit}
\end{figure}


The decreasing strength (in terms of slope of a linear correlation) and statistical significance of
the correlation going from $H\beta$- to Mg~II to C~IV-estimated FWHM black hole masses could be 
due to the different average luminosities of the three subsamples (Fig.~1, 
bottom panel), or could be due to a problem related to the estimates of $M_{BH}$.
We directly explored both possibilities:\\
$\bullet$ {\bf Luminosity effects.} The luminosity intervals of the three subsamples overlap significantly.
 We analyzed the $\Gamma_X$-L/L$_{EDD}$ correlation for luminosity-limited
subsamples of the
Mg~II group, matched to the luminosity distribution of the $H\beta$ group. We performed
Pearson's tests and linear correlation fits for Mg~II objects with bolometric luminosities
log($L_{BOL}$)$<$45.5 (23 objects), log($L_{BOL}$)$<$45.8 (76 objects), and log($L_{BOL}$)$<$46 (123 objects). In no case did we find a better correlation than in the whole Mg~II sample. This suggests that
luminosity is not the main reason for the observed trend.
\\
$\bullet$ {\bf Black hole mass estimates.}
We compared the black hole mass estimates from pairs of lines in the S08 sample. 
As discussed in S08, about 8,000 objects in their sample have black hole mass 
estimates from both $H\beta$ and Mg~II, and about 5,000 from both Mg~II and C~IV.
In Fig.~3 we plot the mass estimates for the two groups (this plot is analogous to Fig.~6 in S08),
with a best fit linear correlation. 
We find a strong (though not linear) correlation between 
H$\beta$ and Mg~II masses, while no significant correlation is found between Mg~II and C~IV masses.
These two findings have two different possible explanations:\\
- Both H$\beta$ and Mg~II are expected to be produced by virialized gas, and are therefore 
in principle good mass indicators. 
The mismatch between H$\beta$ and Mg~II masses is likely related to the uncertainties
in the measurement of the Mg~II line parameters: the line is a doublet, and the
estimate of the strength of the contaminating Fe~II emission is more problematic than for the H$\beta$ line.
Recently Onken \& Kollmeier~(2008) discussed this issue and found
that a correction has to be applied to the average Mg~II based mass distribution, in order to make it compatible with
the one based on H$\beta$. In our case, in order to apply a correction to the Mg~II-based values of $M_{BH}$ 
in single sources,
we used the best fit linear correlation shown in Fig.~3A (log[M$_{BH}$(H$\beta$)]=1.8$\times$log[M$_{BH}$(Mg~II)]-6.8),
under the assumption that the $H\beta$-based masses are correct. 
This should improve the precision of the estimates of $M_{BH}$, at least
in a statistical sense. We repeated the same statistical analysis on the so-modified Mg~II subsample, and we obtained a stronger correlation (probability of null correlation $<10^{-6}$), and a 
linear correlation coefficient r=0.30. This is still lower than that found for $H\beta$ sample,
as expected, since our correction is only statistical, and a significant
residual dispersion is still present between the $H\beta$-based and Mg~II-based estimates 
(Fig.~3, upper panel), However, the strength of the correlation 
is significantly higher than that found with the 
original $M_{BH}$ values (r=0.21, Table~1).
\\
- C~IV is a high ionization line, produced in the inner broad line region by gas probably
having non-virialized components (for example, associated to an outflow). This makes this
line a poor estimator of the black hole mass (Baskin \& Laor~2005, Netzer et al.~2007).
The lack of correlation between Mg~II and C~IV masses confirms this finding, and provides 
an explanation for the absence of correlations between the X-ray spectral slope and the
Eddington ratio and the black hole mass.
  
\section{Conclusions}

We presented a   
strong correlation between the X-ray photon index, $\Gamma_X$, and the Eddington ratio, L/L$_{EDD}$
for a sample of $\sim$400 AGNs having good quality X-ray and optical spectra from the SDSS quasar
survey and serendipitous {\em XMM-Newton} observations.
A weaker, but statistically significant, correlation between $\Gamma_X$ and the black hole mass $M_{BH}$
was also found.

The correlations found here are important in two respects: 
first, a
$\Gamma_X$-L/L$_{EDD}$ correlation is an
important constraint for theoretical emission models. In general, it suggests a strong link
between the accretion rate and the physical conditions in the hot corona producing the X-rays.
For example, 
recently Cao~(2008) showed that a $\Gamma_X$-L/L$_{EDD}$ relation, together with the $\alpha_{OX}$-luminosity
relation, depend on the balance between the magnetic field and the gas and electron pressures
in a disc-corona model where the corona is heated by magnetic field reconnections.
 
Second, as already pointed out by Shemmer et al.~(2008), 
our results show that the X-ray slope can be used as an Eddington ratio estimator. Since the
X-ray luminosity can be converted to a total luminosity through bolometric corrections and the $\alpha_{OX}$-luminosity correlation, an estimate of the black hole mass can be obtained from the
X-ray data alone. Given the
high dispersion of the correlation (0.4 dex), this is not at present a reliable method for single sources.
However, it could be a powerful technique when used to estimate average values in large samples, such
as the ones which are expected to be available in the future from new missions like E-Rosita.

We are currently working at 
expanding SDSS-XMM quasar survey, using the DR7 version of the SDSS\footnote{http://www.sdss.org/DR7/} 
and the 2009 {\em XMM-Newton} archive.
This new sample will
have about twice as many quasars with good quality X-ray spectra, and will allow further investigations 
of the multi-wavelength correlations among AGNs. 
\acknowledgements

We are grateful to the referee for his/her constructive comments.
This work has been partly supported by grants prin-miur 
2006025203, ASI-INAF I/088/06/0, and  NASA  NNX07AI22G.



\end{document}